# Development and Evaluation of a Narrow Linewidth Laser System for $^{171}$Yb$^+$ E2 Transition

Yani Zuo, Shiying Cao, Shaoyang Dai, Yige Lin, Tao Yang, Baike Lin, Fei Meng, Weiliang Chen, Kun Liu, Fasong Zheng, Tianchu Li, and Fang Fang

*Abstract*—We report the construction and characterization of a narrow-linewidth laser system for the interrogation of the E2 clock transitions at 436 nm of ytterbium ions trapped in end-cap traps. The 871 nm seed laser at the fundamental frequency is referenced to a 10 cm long notched ULE cavity. The output of the laser system is delivered to a narrow-linewidth femtosecond fiber comb, which has been referenced to an ultrastable 698 nm laser, with a phase noise-canceled fiber link. The beat between the laser and the comb shows a sub-Hz linewidth, and with a stability better than 2E-15@1~100 s. The performance of the self-developed wavelength extension ports at 871 nm of the narrow linewidth erbium-doped fiber comb with single-point frequency-doubling technique is also verified.

## I. Introduction

With the development of optical frequency comb and ultra-stable laser technologies, optical clocks based on either optical lattice trapped atoms or trapped ions, whose uncertainty exceeds the current SI second definition realization cesium fountain clocks [1][2], are the most promising candidates for metrology advances such as the unit redefinition [3]. And compact optical clocks allow for chronometric leveling between distant locations and many other fundamental physical tests [4].

Among diverse optical clock schemes, the ytterbium ion optical clock has a broad application prospect [5][6][7][8]. 171-isotope ytterbium ion has two clock transitions (E2 and E3) accepted as secondary representations of the second. The E3 transition of 171-isotope ytterbium ion is insensitive to external field perturbations, exhibits strong relativistic effect, and thus has linewidth at the level of nHz. The ytterbium ion has a relatively large atomic mass and the cooling laser wavelength is close to the dissociation wavelength of the dark Hydride molecular ion, so the ytterbium ions have long storage time. Also, lasers for manipulating ytterbium ions are accessible and relatively stable. These advantages make it possible to establish ultra-low uncertainty optical clock, to carry out more precision measurements.

This work was supported in part by the National Key R&D Program of China, National Quality Infrastructure under Grant 2021YFF0603800 and in part by the Fundamental Research Funds of National Institute of Metrology of China under Grant AKYCX2206. (Corresponding author: Fang Fang)

Cavity-stabilized narrow linewidth lasers [9][10] are the local oscillators of optical clocks, whose linewidth and frequency stability directly affect the performance of the optical clock. And they also have a variety of applications such as photonic generation of low-phase-noise microwave, gravitational wave detection, and ultra-coherence spectroscopy.

We firstly build the Yb$^+$ optical clock with the quadrupole transition (E2) as the clock transition to verify the feasibility of the system. It is worth noting that the relatively short lifetime (53 ms) will limit the final performance of the E2 transition ytterbium ion clock, which leads to a compromise of the system design. The ytterbium ion optical clock system includes the optical system, the physical system, and the data acquisition system. The system diagram is shown in Fig. 1.

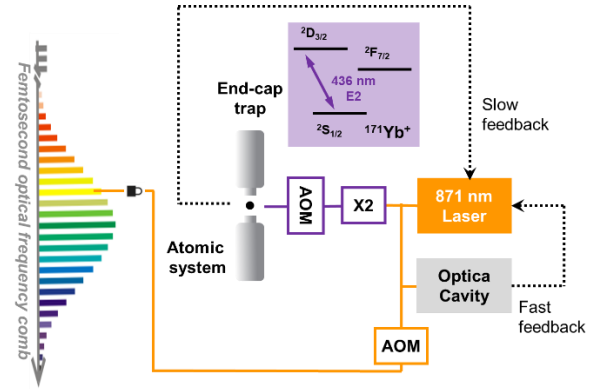

**Fig. 1.** The scheme diagram of the $^{171}$Yb$^+$ E2 optical clock.

We present and characterize an 871 nm laser system applied for the E2 transition interrogation following the Conference on Precision Electromagnetic Measurements (CPEM) proceedings article[11] where we primarily demonstrate the system's design. An external cavity diode laser (ECDL) at 871 nm was frequency stabilized to a high-finesse optical cavity with Pound-Drever-Hall (PDH) method before second harmonic generation (SHG). This extended article reports the laser system characterization results beyond the conference version. Since only one set of 871nm laser system was built, in order to evaluate the performance of the laser system, a narrow-linewidth optical frequency comb at another laboratory was modified for

The authors are with the Division of Time and Frequency Metrology, National Institute of Metrology (NIM), Beijing 100029, China (e-mail: zuoiyn@nim.ac.cn; fangf@nim.ac.cn).

beat note measurement. The comb is referenced to a 30-cm-long cavity-stabilized 698 nm laser system serving a strontium lattice clock.

This work also established two phase-noise-canceled fiber links to deliver the laser to the reference cavity and the evaluation beat-note module, effectively limiting the additional frequency noise introduced by the reference system. The performance of the 871 nm wavelength extension port of the narrow linewidth erbium-doped fiber comb using single-point frequency-doubling technique was constructed and verified. Based on the evaluation system, this paper demonstrates the stability and linewidth of the ultra-stable laser system, meeting the current requirements of the E2 $Yb^+$ optical clock.

## II. SETUP OF THE INTERROGATION LASER SYSTEM

The maximum excitation probability of a clock transition and the short-term stability of an optical clock will benefit significantly with the improvement of the interrogation laser stability. Considering the SNR of the ion optical clock and the natural linewidth of the E2 transition, we expect the short-term stability of the local oscillator to be below the system QPN stability limit[13][14]. A 10-cm cavity scheme was used in this paper, there have been many studies on this length cavity frequency-stabilized laser[17]-[27].

Therefore, a modified version of the 10-cm cavity design was adopted[11], shown in fig. 2. The cavity spacer's diameter is 5 cm with a horizontal notch structure. The cavity is directly supported by 4 fluorinated rubber Viton pads. The support position of the cavity is optimized by the FEA method. The acceleration sensitivity of the cavity is estimated to be $4E-10/g$ along the relevant direction[15][16].

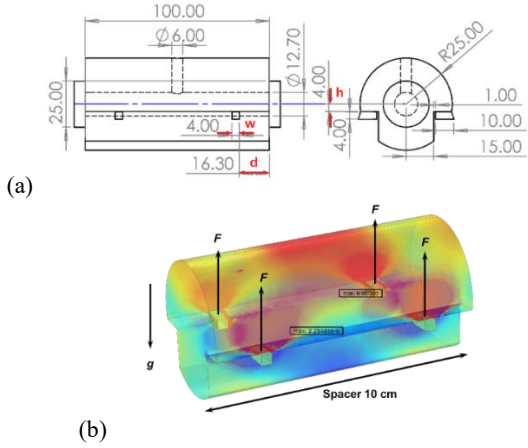

(a)

(b)

**Fig. 2.** Drawing of the optical cavity. (a) shows the optical cavity is support by four Viton pads. Dimensions are in mm. (b) shows the spacer displacement under acceleration insensitive holding position in Finite element analysis (FEA).

A notched cylindrical optical cavity is held in a vacuum chamber providing effective isolation from environmental disturbances, as shown in fig. 3. The cavity spacer is made of ULE glass with a low expansion coefficient. To control the temperature of the vacuum chamber, an active temperature control layer is combined with two inner passive temperature shields. A pair of high-reflectivity flat-concave mirrors are optically contacted to the ULE spacer. The radius of curvature of the concave mirror is 50 cm, and the cavity mirror adopts FS substrates. As estimated, the thermal noise limit of this reference cavity is about $5E-16$[12]. The vacuum chamber uses a 20 l/s ion pump to maintain a pressure at $1E-5$ Pa. In order to isolate the influence of the ambient vibration, the vacuum chamber is placed on an active isolation platform.

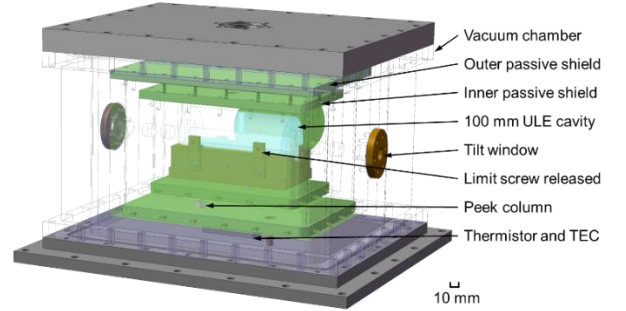

**Fig. 3**. Drawing of the cavity mounted on vacuum chamber.

The optical layout diagram of the cavity-stabled laser system is shown in Fig. 4.

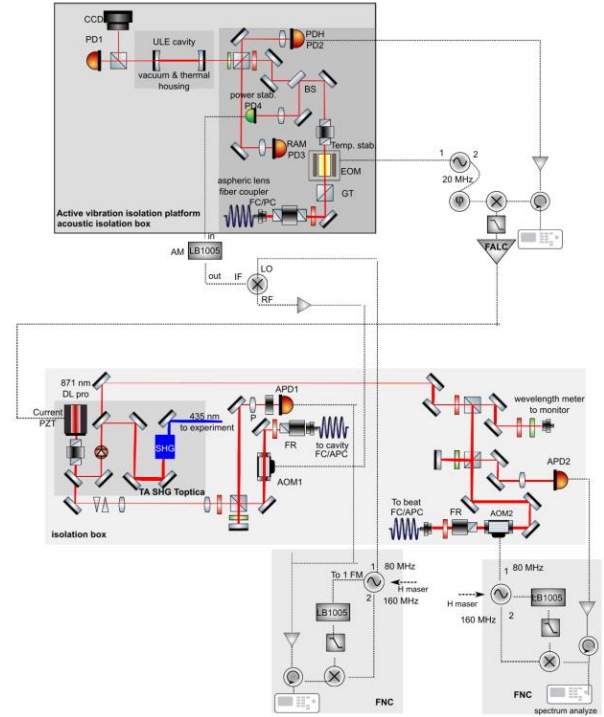

**Fig. 4**. Schematic diagram of the cavity-stabilized system. Red lines indicate optical signals, black lines denote the propagation of the electronic signal. All the components are height fixed at 50 mm to reduce vibration sensitivity. The dark gray block represents the optical breadboard placed on an active vibration isolation platform. FR: Faraday rotator; EOM: electro-optic modulator; AOM: acousto-optic modulator; PD: photo detector; APD: avalanche photodiode; FNC: fiber noise cancellation; SMPM: single mode polarization maintaining.

The laser is a commercial ECDL laser, where a seed laser at 871 nm is amplified and frequency-doubled to 436 nm. The laser light was guided through a 5 m long single-mode polarization-maintaining (SMPM) fiber to an active vibration-isolation platform (AVI-200) in a homemade acoustic isolation. The output end of this fiber is flat, so a small part of the light reflects back for fiber noise cancellation. And the laser head of the 871 nm Toptica DL-Pro ECDL has a DC-coupled (DC Mod.) current modulation port, which can be used as a high-speed feedback frequency control port.

The modulation frequency of the EOM is 20 MHz. In order to further reduce the influence of the RAM in the EOM [28][29][30], the polarization and incident direction of the light is carefully adjusted, and two 60 dB and 35 dB isolations are placed before and after the EOM. A fast low-noise photodetector detects the error signal, and the signal is demodulated and low-pass filtered to obtain a frequency discrimination signal. The laser frequency is corrected by adjusting the laser current and PZT. The locking bandwidth of the system is about 1.5 MHz. To reduce beam distortion and improve alignment stability, the optical height is fixed to 50 cm above the breadboard surface and many homemade support mounts without flexible structures like springs were adopted. Through the cavity ring-down method, the fineness of the cavity is measured to be about 200,000.

Fiber noise cancellation (FNC) is employed to suppress the phase noise induced by the optical fiber that transmits light from the optical bench to the optical cavity. The intensity of the light reflected by the polarization beam splitter (PBS) before the optical cavity is stabilized by adjusting the driving power of the acoustic-optic modulator (AOM) used for the fiber noise cancellation, as shown in Fig 5.

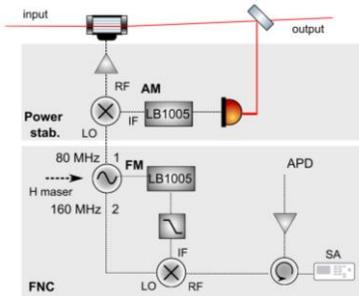

**Fig. 5**. Diagram of the multifunctional control of the AOM.

Before collimated into the SMPM, the light is frequency shifted by a +1-order single pass AOM working at 80 MHz. Henceforth, the AOM will play a multifunctional role in the control system, such as cavity drift compensation, optical fiber noise cancellation, and power stabilization.

### III. CHARACTERIZATION OF THE LASER SYSTEM

Since only one 871 nm laser was built, for cost reduction and quick evaluation, a narrow-linewidth optical frequency comb referenced[32] to the 30-cm-long cavity-stabilized 698 nm laser system was used to perform the beat-note measurement. Since the stability of the 698 nm laser is at the order of E-16[31], the stability of the beat signal reflects the stability of the 871 nm laser system.

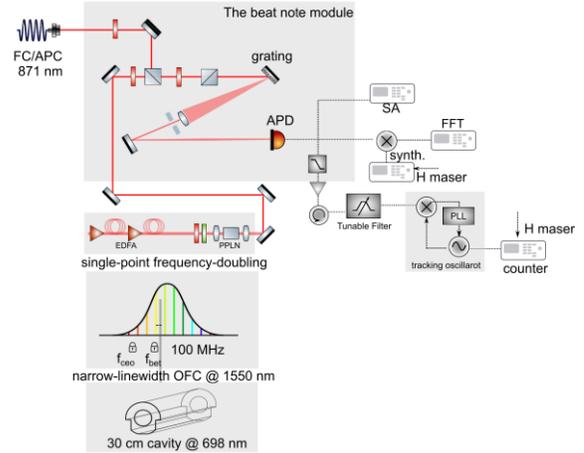

**Fig. 6.** Scheme of the beat-note measurement based on a narrow-linewidth optical frequency comb (OFC) referenced to the 30 cm-long-cavity stabilized 698 nm laser system.

A fiber link with a length around 500 m between these two systems was established. The OFC can faithfully transfer the coherence from the 698 nm reference laser to any wavelength ranging from visible to infrared by single-point frequency-doubling technique. The 871 nm laser spatial overlap with one of the comb branches after amplification, single-point frequency-doubling, and filtering. The heterodyne beat is down-converted and measured by a frequency counter. The beat note signal was filtered by a tracking filter in order to clean up the noise and regenerate the signal-to-noise ratio (SNR). The bandwidth of the tracking was adjusted to avoid copying the electronic noise.

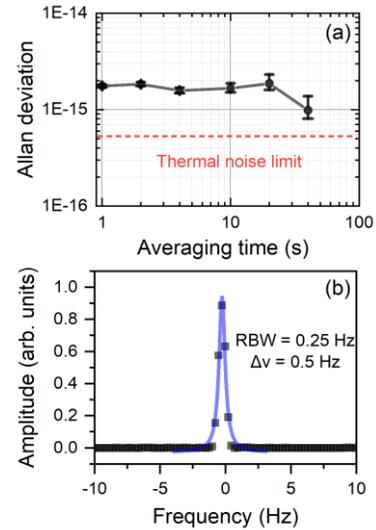

**Fig. 7.** Laser stability and line-width measurements. Panel (a) shows the fractional frequency instability in ADEV of the cavity-stabilized 871 nm laser system along with its calculated thermal noise floor. Panel (b) shows the spectrum of beat note reaches sub-Hz level when the RBW of 0.25 Hz.

The beat exhibits a linear drift on the order of 2 Hz/s. Figure 7 shows a 5-min time series of the beat note with this linear drift removed. Allan deviations are computed and shown in Fig. 7(a). The frequency instability of the heterodyne beat is reduced to $1.4\times10^{-15}$ around 1 s. Currently, two noise contributors, the vibration noise and the temperature fluctuation of the passive thermal chamber, limit the instability of the laser to be further reduced down to the thermal-noise limited value of $5\times10^{-16}$. The linewidth of the beat note between the laser and the comb is 0.5 Hz.

## IV. Conclusion and Discussion

We have developed an ultra-stable 871-nm laser referenced to a 10-cm optical cavity operated at room temperature. The frequency instability of the cavity-stabilized 871 nm laser reaches a minimum of 1.4E-15 around 1 s. The measured linewidth of the cavity-stabilized laser is 0.5 Hz. The comparison between different wavelengths (698 nm, 871 nm) was achieved using a phase-coherent fiber link. This also verified the measurement capability of the narrow linewidth OFC at this wavelength. These results laid the foundation for ongoing projects on the ytterbium optical ion clock. The evaluation system enables the future comparison of two optical frequency ratio measurements. The performance of the laser system is good enough to be the interrogation laser for the E2 transition. Further improvements will be considered to enhance cavity-temperature control and optimize the vibration-insensitive cavity-support structure. And more precision comparison between different frequencies should also eliminate inter-branch non-common-mode noise involved in OFCs[33][34].